# Specific Ion Effects of Trivalent Cations on the Structure and Charging State of β-Lactoglobulin Adsorption Layers

*Manuela E. Richert, Georgi G. Gochev, Björn Braunschweig*[*]

*Institute of Physical Chemistry and Center for Soft Nanoscience, Westfälische Wilhelms-Universität* Münster Corrensstraße 28/30, 48149 Münster, Germany

\* Corresponding author e-mail: braunschweig@uni-muenster.de

**ABSTRACT**

Properties of proteins at interfaces are important to a many processes as well as in soft matter materials such as aqueous foam. Particularly, the protein interfacial behavior is strongly linked to different factors like the solution pH or the presence of electrolytes. Here the nature of the electrolyte ions can significantly modify the interfacial properties of proteins. Therefore, molecular level studies on interfacial structures and charging states are needed. In this work, we addressed the effects of $Y^{3+}$ and $Nd^{3+}$ cations on the adsorption of the whey protein β-lactoglobulin (BLG) at air-water interfaces as a function of electrolyte concentration. Both cations caused very similar but dramatic changes at the interface and in the bulk solution. Here, measurements of the electrophoretic mobility and vibrational sum-frequency generation spectroscopy (SFG) were applied and consistently showed a reversal of the BLG net charge at remarkably low ion concentrations of 30 (bulk) and 40 (interface) µM of $Y^{3+}$ or $Nd^{3+}$ for a BLG concentration of 15 µM. SFG spectra of carboxylate stretching vibrations from Asp or Glu residues of interfacial BLG showed significant changes in the carboxylate stretching frequency, which we associate to specific and efficient binding of $Y^{3+}$ or $Nd^{3+}$ ions to the proteins carboxylate groups. Characteristic reentrant condensation for BLG moieties with bound trivalent ions was found in a broad concentration range around the point of zero net charge. The highest colloidal stability of BLG was found for ion concentrations <20 µM and >50 µM. Investigations on macroscopic foams from BLG solutions, revealed the existence of structure-property relations between the interfacial charging state and the foam stability. In fact, a minimum in foam stability at 20 µM ion concentration was found when the interfacial net charge was negligible. At this concentration, we propose that the persistent BLG molecules and weakly charged BLG aggregates drive foam stability, while outside the bulk reentrant zone the electrostatic disjoining pressure inside foam lamellae dominates foam stability. Our results provide new information on the charge reversal at the liquid-gas interface of protein/ion dispersions. Therefore, we see our findings as an important step in the clarification of reentrant condensation effects at interfaces and their relevance to foam stability.

**INTRODUCTION**

Proteins are of significant interest because of their biological functions but also due to their application as stabilizers[1] in food products, cosmetics and pharmaceuticals. Interactions of trivalent cations with proteins play an important role in physiological and biological processes like in the signal transduction.[2] Furthermore, lanthanides were found to be cofactors for enzymes in methylotrophic bacteria where they can strongly influence the protein conformation.[3]

Different from biochemical aspects, the interaction of trivalent ions with proteins and surfactants has gained considerable attention in case of protein crystallization[4–6] but also for sensing purposes that are able to detect trace amounts of rare earth elements[7] or for ion flotation.[8,9] Efficient binding of e.g. $Nd^{3+}$ or $La^{3+}$ ions to interface-adsorbed proteins in foams and emulsions is also interesting for ion extraction from liquids and may evolve as an important strategy to recycle rare earth elements.[10–12]

In fact, proteins are known to adsorb at fluid interfaces that are ubiquitous in aqueous foams and emulsions where the proteins form viscoelastic adsorbate layers.[13–15] The surface activity of proteins, their surface rheology and interfacial charging state are strongly linked to the solvent conditions and significantly vary with pH and ionic strength.[15–18] The latter determine protein-protein interactions as well as their aggregation behavior in the bulk,[19] and ultimately the properties of surface-adsorbed proteins. Not only the ionic strength, but also the chemical nature and especially the valence of ions can greatly effect bulk[4,6,20–2324] and interfacial[6,18,23,25] properties of proteins.

β-lactoglobulin (BLG) has been frequently investigated with regard to its behavior in bulk[26–30] solutions and at interfaces.[14–18,23,31–33] and has an isoelectric point (IEP) at pH ≈ 5.1.[31,34] Solid-liquid and liquid-liquid coexisting phases have been identified in the non-equilibrium phase diagram for BLG solutions at the IEP and for weakly contracted Na-acetate /NaCl buffer solution with ionic strengths < 20 mM.[30,35] However, at NaCl concentrations >1 M such solutions turn transparent.[30,36]

Very similar phase behavior is observed in native aqueous solutions of BLG (pH > IEP) in the presence of trivalent cations.[37–39] Such cations can induce a charge reversal of the protein net charge and cause reentrant condensation. The latter is defined by a phase-separation region (regime II) in the proteins' non-equilibrium phase diagram and is caused by the formation of protein aggregates around the isoelectric state.[37–39] At concentration below and above the phase separation region, the solutions are clear (regimes I and III).[4] For globular proteins such as BLG, ovalbumin or serum albumins, reentrant condensation has been observed in in the presence of $Y^{3+}$, $Fe^{3+}$, $La^{3+}$ and $Al^{3+}$.[37–40] At this point one should note that hydrolysis of metal salts can significantly modify the pH of aqueous solutions and that the extent of this modification is directly dependent on the used salt.[37,39] For instance, changes in the protein solution pH are pronounced for $Fe^{3+}$ and are nearly absent for $Y^{3+}$. In case of $Fe^{3+}$, the pH distinctly decreases with increasing salt concentration.[39] Consequently, the reentrant regime occurs at lower concentrations for $Fe^{3+}$ and is much narrower in terms of cation concentrations as compared to $Y^{3+}$.[37,39] Such coupling of pH and the effects of trivalent ions makes the description of the protein charging state much more complex. In this study, we, therefore work with $Nd^{3+}$ and $Y^{3+}$ ions in order to minimize possible pH changes.

Nevertheless, previous studies have clearly shown that the protein isoelectric state in bulk solutions can be reached either by pH adjustment or by adding trivalent cations. In the latter case, the mechanism of the protein charge inversion has been attributed to electrostatically and entropy driven binding of trivalent cations to the protein.[39,41,42] Further, it was proposed that the binding of trivalent ions takes place via ion chelation at carboxylate groups of Asp and Glu amino acid residues.[39,42] Roosen-Runge et al.[43] described the interactions between proteins and trivalent cations in a way that proteins are modeled as patchy particles, while multivalent cations are handled as a bridging species. Simultaneous binding of such cations to patches of different protein molecules can induce binding between these molecules, which is considered as a main driving force for protein clustering and aggregation in the reentrant condensation regime. Using molecular dynamics simulations, Beierlein et al.[18] showed the influence of different monovalent cations on BLG monomers and dimers. They also proposed bridging by $Li^+$ cations and ion pairing with carboxylate groups from Asp and Glu residues at the BLG surface.[18]

In fact, a number of properties of BLG at soft fluid interfaces have been examined as dependent on pH as well as on the concentration of monovalent or divalent metal salts.[15,17,18,23,31,44,45] In case of $Ca^{2+}$, this was done with sum-frequency generation (SFG) spectroscopy, which showed a charge reversal at the interface that is accompanied by an increase of the thickness of interfacial layers and an increase of the foam stability.[23] This is similar to what Sarker et al.[25] showed for $Al^{3+}$ ions, where an improved foamability and foam stability was reported for BLG and Tween 20. However, in this early work, BLG/non-ionic surfactant mixtures were investigated and therefore reentrant effects could not be solely identified. This was done in the very recent study of Fries et al.[6] for bovine serum albumin/$Y^{3+}$ solutions, where the authors demonstrated that reentrant effects in the bulk are reflected at solid-liquid interfaces. However, information on reentrant effects at other interfaces e.g. liquid-gas interfaces is missing so far.

The current knowledge on the influence of trivalent ions on the interfacial properties of proteins is mainly based only on the works by Fries et al.[6] and Sarker et al.[25] As mentioned above, the latter work was done with BLG solutions containing a non-ionic surfactant and additionally with a cation that is known to cause pH effects. Therefore, the pure individual influence of the cations remained largely unclear. The aim of the present study is to close this gap. Here, we provide new molecular-level information on the interfacial properties of the model protein BLG as a function of $NdCl_3$ and $YCl_3$ concentrations. We have applied a multi-technique approach that addresses different length scales of aqueous foam – from BLG molecules and their interaction with ions in the bulk to the air-water interfaces and macroscopic foam.

**EXPERIMENTAL DETAILS**

**Sample preparation**

β-Lactoglobulin (BLG) was isolated and purified (≥99%) from whey protein isolate,[46] and was kindly provided by the Kulozik group (TU München, Germany). $YCl_3$ (≥99.99%) and $NdCl_3·6H_2O$ (≥99.9%) were both purchased from Sigma Aldrich (now Merck) and were used as received. Both salts were dissolved in nitrogen flushed ultrapure water from a Millipore ReferenceA+ purification system (18.2 MΩcm; total organic carbon content <5 ppb). BLG was also solved in purified water. For that the samples where gently shaken and were left to equilibrate for at least one hour. Stock solutions of the salts were prepared with concentrations of 1 and 10 mM. In all experiments, BLG concentrations were fixed to 15 µM (~0.27 mg/ml) and the salt concentration was varied between 2 and 500 µM. At least 20 min before measurements, appropriate aliquots of $YCl_3$ or $NdCl_3$ electrolytes were added to the protein solution. Prior to loading of a sample for measurement, the solution was gently shaken to reach homogeneity. This was especially important for solutions where the salt concentration led to the formation of larger protein aggregates as it will be discussed below. The used glassware for sample preparation and measurements was socked for at least 12 h in a mixture of concentrated sulfuric acid (98 %, Rotipuran, Carl Roth) and Nochromix (Godax Labs) and was subsequently exhaustively rinsed with ultrapure water afterwards.

**Electrophoretic mobility ($\mu_\zeta$)**

Measurements of the electrophoretic mobility were performed using a Zetasizer ZSP (Malvern, UK). Each sample was measured at least 3 times and mean values and standard deviations were calculated. We decided not to convert $\mu_\zeta$ into ζ-potentials as this would require to go from the Hückel to the Helmoltz-Smulochowski regime as the protein (aggregate) size and the ionic strength is expected to change significantly (see below).

**Optical Density (OD)**

The OD of all samples was recorded with a Lambda 650 (Perkin Elmer, USA) UV-VIS spectrometer and was evaluated at a fixed wavelength of 410 nm where no significant bands from pure $NdCl_3$ solutions were observed. To study the sedimentation of turbid samples in the phase separation regime, measurements were performed directly after adding salt to the protein solution (Figure 1a) as well as after 1 and 4 h (Supporting Information, Figure S1). This was done by taking samples from the supernatant.

**Sum Frequency Generation (SFG)**

SFG spectroscopy is a second-order nonlinear optical method that is inherently surface specific for materials with inversion symmetry such as isotropic liquids and gases.[16] In order to generate sum-frequency (SF) signals, two laser beams - a visible (VIS) and a tunable broadband infrared (IR) - with frequencies $\omega_{vis}$ and $\omega_{IR}$ are overlapped in space and time at the air-water interface. The intensity of the resulting SF signal is dependent on contributions from second- and third-order electric susceptibilities $\chi^{(2)}$ and $\chi^{(3)}$:[47,48]

$$I(\omega_{SF}) \propto \left| \chi^{(2)} + \frac{\kappa}{\kappa + i\Delta k_z} \chi^{(3)} \phi_0 \right|^2 \quad (1)$$

$\chi^{(2)}$ results directly from the interface and is comprised by non-resonant $\chi^{(2)}_{NR}$ and resonant contributions:

$$\chi^{(2)} = \chi^{(2)}_{NR} + \sum_k \frac{A_k}{\omega_k - \omega_{IR} + i\Gamma_k} \quad (2)$$

The oscillator strength $A_k = N \int f(\Omega) \beta_k(\Omega) \, d\Omega$, which is represented by the amplitude of a vibrational mode k, depends on the number density $N$ of interface-adsorbed molecules, their orientational distribution f($\Omega$) and on their hyperpolarizability $\beta_k$. Depending on the net orientation of molecules at an interface, the amplitude $A_k$ can have different polarities (or phases) which determine the shape of a vibrational band in SFG spectra that is caused by its coherent overlap with other vibrational bands according to equations (1) and (2). $\chi^{(3)}$ is relevant in the presence of an electric double layer with the double-layer potential $\phi_0$. For that reason, not only information on the interfacial composition from existing vibrational bands can be gained from SFG, but also qualitative and, in some cases, also quantitative information on the interfacial charging state.[47–51]

For SFG spectroscopy, we used the Münster Ultra-Fast Spectrometer for Interface Chemistry (MUSIC), which has been described in detail elsewhere.[47] In brief, we used a SolsticeAce amplifier system (Spectra Physics, USA) with TopasPrime and NDFG units (Light Conversion, Lithuania) to generate a tunable femtosecond IR beam with a bandwidth >300 cm$^{-1}$ and a narrowband (4 cm$^{-1}$) visible picosecond (ps) beam at a fixed wavelength of 804.1 nm. Both laser beams were spatially and temporally overlapped at the interface of interest and the reflected sum frequency photons were collected and directed to a spectrograph (Andor Kymera) with an EMCCD (Andor Newton). After 30 min equilibration time, SFG spectra from BLG modified air-water interfaces were recorded in ssp polarizations (s: SF-, s: VIS- and p: IR-beam) and were normalized to the SFG spectrum of an air-plasma cleaned thin Au film. SFG spectra in the frequency region from 2800 to 3800 cm$^{-1}$ were collected in five scanning steps and with an acquisition time of at least 60 s per IR wavelength. Additional spectra were taken in the amid region from 1300 to 1750 cm$^{-1}$ which was done by scanning the IR frequency in three steps with an acquisition time of 120 s per IR wavelength.

**Surface Tensiometry**

Dynamic surface tensions γ(*t*) were recorded with the pendant drop technique using DSA 100 (Krüss, Germany) and PAT-1 (Sinterface, Germany) tensiometers. The surface tension was taken after an adsorption time of 30 min and averaged from several measurements.

**Foam Stability**

Analysis of macroscopic foams was performed with a DFA 100 (Krüss, Germany) foam analyzer, with which the liquid and foam heights were recorded optically. Foams were generated in a glass column with a diameter of 4 cm. For foam formation, ambient air was purged with a gas flow of 0.3 Lmin$^{-1}$ through a porous glass frit for 30 s. The frit had pore sizes from 16 to 40 µm. After the foaming process was completed the initial height $h_{max}$ was measured and, subsequently, the loss in foam height was monitored as a function of foam age *t*. This procedure was repeated for different salt concentrations. In the discussion below, we present results on foam stability (FS), which we define here as follows: $FS = (h_{30min}/h_{max}) \cdot 100$. Here, $h_{max}$ is the initial foam height after the foaming process was completed and $h_{30min}$ is the foam height after a waiting time of 30 min.

For trivalent cation concentrations of >100 µM, the foam experiments were impaired by foam residues that stuck at the column walls and led to an unsatisfactory quality of the foam height detection by the Krüss instrument. Possibly this was caused by electrostatic interactions of the positively charged proteins (see discussion) with the negatively charged silica surfaces of the glass column. For that reason, ion concentrations >100 µM were omitted in our analysis.

## RESULTS and DISCUSSION

### Bulk charging state and phase separation

Figure 1a presents the electrophoretic mobility as a function of electrolyte concentration for both, $NdCl_3$ and $YCl_3$, electrolytes and the samples' optical density (OD). A close inspection of Figure 1a clearly demonstrates that the electrophoretic mobility in the bulk and the samples' OD show a remarkable dependence on the concentration of $Nd^{3+}$ and $Y^{3+}$ ions. In absence of any salt and at $YCl_3$ or $NdCl_3$ concentrations <5 µM, which corresponds to ion/protein molar ratios $c_{ion}/c_p$ of <1/3, negative electrophoretic mobilities were observed. Increasing the $YCl_3$ or $NdCl_3$ concentrations has caused a considerable increase in the electrophoretic mobility until the latter approached negligible values at 30 µM ($c_i/c_p \sim 2$) and increased further to a plateau value at the highest $YCl_3$ and $NdCl_3$ concentrations of 500 µM.

Because the isoelectric point (IEP) of BLG is found at pH ~5.1[31] and the pH of 15 µM BLG in water is ~6.5, we point out that, at our experimental conditions, BLG molecules in the bulk solution carry a negative net charge. According to previous titration studies,[52,53] the BLG net charge at this pH is about -8$e$, which is consistent with negative electrophoretic mobilities. On the other hand, at solution pH ≥ 6, hydrolysis should be taken into an account. Under the slightly acidic conditions in our study only the first hydrolysis step (with equilibrium constant of $pK_0 \approx 7.7$) can have, however small, contribution.[39]

Furthermore, the speciation of multivalent ions could be modified in presence of acidic proteins. To the best of our knowledge, there are no literature information about association of trivalent cations to protein side chains. Roosen-Runge et al.[39] proposed an analytical model for the association of trivalent cations to the protein surface and estimated an association constant $pK_m \approx 4.2$ for a cation and one exposed carboxylic group. The authors also assumed that no induced speciation of the trivalent cation hydrolysis products occur due to such interaction. Therefore, we assume that the metal ions in the protein solutions of our study are predominantly in their trivalent form.

We now define the concentration where the electrophoretic mobility reaches zero values as the point of zero net charge (PZC). Previous works[18,23] on the interaction of BLG molecules with $CaCl_2$ and LiCl electrolytes have shown that substantially higher $c_i/c_p$ molar ratios of >10 000 were needed in order reach the PZC.

Obviously, the binding of $Nd^{3+}$ and $Y^{3+}$ ions to BLG molecules is orders of magnitudes more efficient as compared to divalent or monovalent cations.[18,23] Furthermore, the latter studies with monovalent and divalent cations were also performed with chloride anions. For that reason, we can conclude that the substantial changes in Figure 1a are cation specific and must be attributed to the presence of $Nd^{3+}$ and $Y^{3+}$.

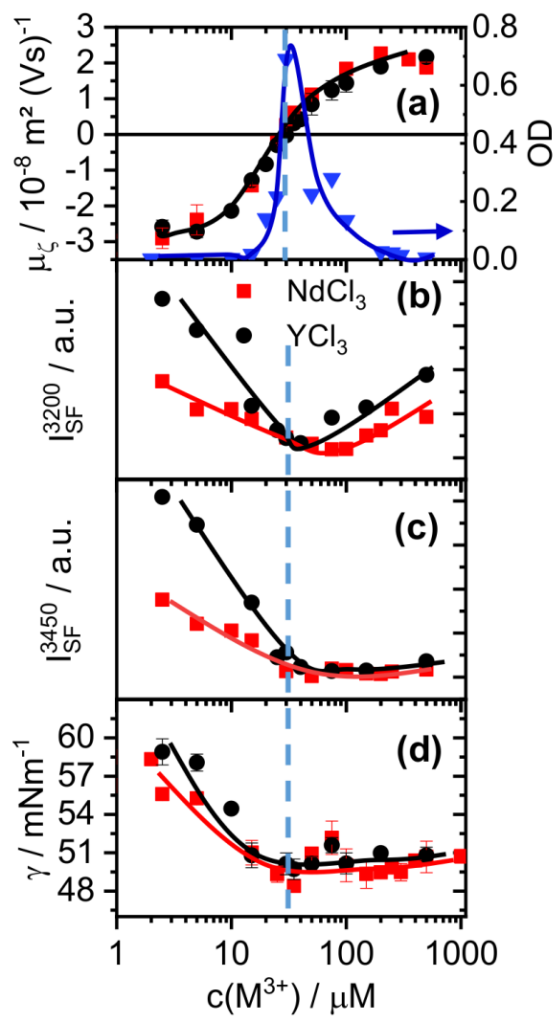

**Figure 1** *Effects of NdCl$_3$ (red squares) and YCl$_3$ (black circles) electrolytes (c(M$^{3+}$)) with 15 µM β-lactoglobulin (BLG) on (a) the electrophoretic mobility µ$_\zeta$ of BLG molecules in the bulk solution and the solutions' optical density (OD) at 410 nm (blue triangles). The OD was measured immediately after sample preparation with NdCl$_3$ while the OD after additional waiting time is presented in the Supporting Information. (b) and (c) show SFG intensities of O-H stretching vibrations that were averaged in the frequency range from 3150 to 3300 cm$^{-1}$ ($I_{SF}^{3200}$) and from 3400 to 3550 cm$^{-1}$ ($I_{SF}^{3450}$) as a function of c(M$^{3+}$). (d) Surface tensions γ after 30 min equilibration time in dependence of c(M$^{3+}$). Solid and dashed lines guide the eye.*

The changes in electrophoretic mobility at ~30 µM are accompanied by a substantial increase in the solutions' OD (Figure 1a) with a maximum at ~35 µM, while solution with <5 and >300 µM salt appeared clear and with negligible OD (Figure 1a). The raise in turbidity or OD in the presence of 30 µM trivalent ions is caused by considerable aggregation of BLG proteins and is directly linked to a low colloidal stability and phase separation at these cation/protein concentration ratios.[21,40] In the Supporting Information we provide additional results on the long term stability and sedimentation behavior, however, as this is out of the scope of our present study we concentrate here on the origin of the changes in Figure 1a and their consequences on interfacial properties of BLG and consequently on the foam stability.

With regard to Figure 1a, it is obvious that the bulk aggregation is directly related to the proteins charging state, which can be divided in three concentration regimes. In regime I, BLG molecules carry sufficient net charge to be colloidal stable, while in regime II this net charge is considerably reduced to a point where aggregation and phase separation can take place.[21,40] In regime III, BLG molecules are considerably overcharged and the BLG dispersion becomes colloidal stable again. Previously, such bulk behavior of acidic protein like BLG has been described by Schreiber and co-workers[6,20,21,37–41] as reentrant condensation, where regime II was defined by threshold ion concentrations which were termed as $c_i^*$ and $c_i^{**}$.[37–39] $c_i^*$ marks the onset of phase separation and $c_i^{**}$ indicates the end of the reentrant zone, where the protein aggregates re-dissolve and the solutions become clear again. Zhang et al.[37] established a linear relationship between $c_i^* = c_1 + m^* c_p$ and $c_i^{**} = c_2 + m^{**} c_p$ and the protein concentration $c_p$. Here, $c_1$ and $c_2$ are constants with the unit of concentration. $m^*$ and $m^{**}$ are the binding numbers of ions which are condensed on the protein surface at concentrations of $c_i^*$ and $c_i^{**}$, respectively. For the behavior of BLG in $YCl_3$ electrolytes, Zhang et al.[37] reported the following parameters: $c_1$ = 50 ±20 µM, $c_2$ = 80 ±20 µM, $m^*$ = 0.5 ±0.2 and $m^{**}$ = 3 ±1. Thus, in our experiments with a fixed BLG concentration of 15 µM, we expect $c_i^*$ and $c_i^{**}$ to be 57.5 ±23 µM and 125 ±35 µM, respectively. Taking into account the standard deviations of the reported parameters, the regime II for our system can be predicted to span 34.5 to 160 µM for the concentration of $Y^{3+}$. ~35 µM is consistent with maximum in OD, while at 160 µM the optical density is nearly zero again (Figure 1a). Note that the OD qualitatively reflects number concentration and aggregate size. However, we point out that the experimental data for the OD in Figure 1a represent a non-equilibrium bulk state as both, sedimentation and aggregation behavior, needs to be investigated for a prolonged period of time in order to get a good estimate. From the data, that we present in Supporting Information (Figures S1 and S2) we estimate the range for $c_i^{**}$ to be in between ion concentrations of 120 to 300 µM. Furthermore, it is interesting to note that BLG enters the reentrant zone when only one $Y^{3+}$ or $Nd^{3+}$ ion is bound to one BLG monomer.

**Interfacial behavior of BLG as a function of $Nd^{3+}$ and $Y^{3+}$ concentration**

Figure 1d shows surface tensions of BLG modified air-water interfaces as a function of $Y^{3+}$ and $Nd^{3+}$ ion concentrations. For that the surface tension was taken after 30 min of adsorption time even though the steady-state surface tension for similar BLG concentrations are usually reached at much longer adsorption times.[17] However, we show in Figure S3 of the Supporting Information that the most significant changes in the dynamic surface tension already take place within the first minutes of adsorption and that the trends for the different salt concentrations at 30 min are comparable to that at longer adsorption times. Increasing the salt concentration causes a remarkable reduction in surface tension until a plateau is reached at concentrations >30 µM. For the salt-free BLG solutions, the surface tension of 61.3 mNm$^{-1}$ ± 0.5 mNm$^{-1}$ (not shown in Figure 1d) is in excellent agreement with earlier measurements for an air bubble in solution.[17] The decrease in surface tension can be attributed to a screening effect due to an increase in ionic strength of the solution, which leads to an increase of the surface activity and thus to the surface excess of BLG.[17]

However, the observed substantial decrease in surface tension is too extensive in order to be explained by simple double layer screening at the very low ionic strengths.

In fact, the latter observation points to changes in the interfacial net charge, similar to what we already discussed for the charging state in the bulk solution. However, at an extended charged interface both, local ion concentrations and molecular structures within the corresponding electric double layer, can be different from the bulk.[23]

In order to gain more information on the latter, we applied vibrational SFG spectroscopy. SFG spectra from air-water interfaces with BLG adsorbate layers are presented in Figure 2 as a function of NdCl$_3$ and YCl$_3$ concentration. At low salt concentration, the SFG spectra are dominated by broad O-H stretching bands, which are centered at 3200 and 3450 cm$^{-1}$. These bands are attributable to interfacial water molecules that are hydrogen bonded.[49] Low and high frequency branches of the O-H spectrum correspond to tetrahedrally coordinated "ice like" and non-tetrahedrally coordinated "liquid like" water molecules at the interface.[47–49] In addition, strong narrow bands due to CH$_3$ symmetric stretching vibrations and the CH$_3$ fermi resonance are centered at 2875 and 2936 cm$^{-1}$, respectively. Further, a weak vibrational band that is characteristic for C-H stretching vibrations from aromatic amino acid residues, such as Phe, Tyr or Trp, of interfacial BLG is noticeable at ~3060 cm$^{-1}$ (shaded areas in Figure 2).

Figure 2 clearly demonstrates that both, the intensity of O-H stretching bands as well as the shape of the aromatic C-H stretching band at 3060 cm$^{-1}$, are strongly depended on the concentration of Nd$^{3+}$ and Y$^{3+}$ ions. O-H stretching bands first decrease considerably in SFG intensity and reach very low values until they start to recover and re-increase with growing Nd$^{3+}$ and Y$^{3+}$ concentrations. These changes in O-H intensities can be qualitatively related to the changes in the interfacial charging state according to equation (1) and the apparent double-layer potential $\phi_0$.

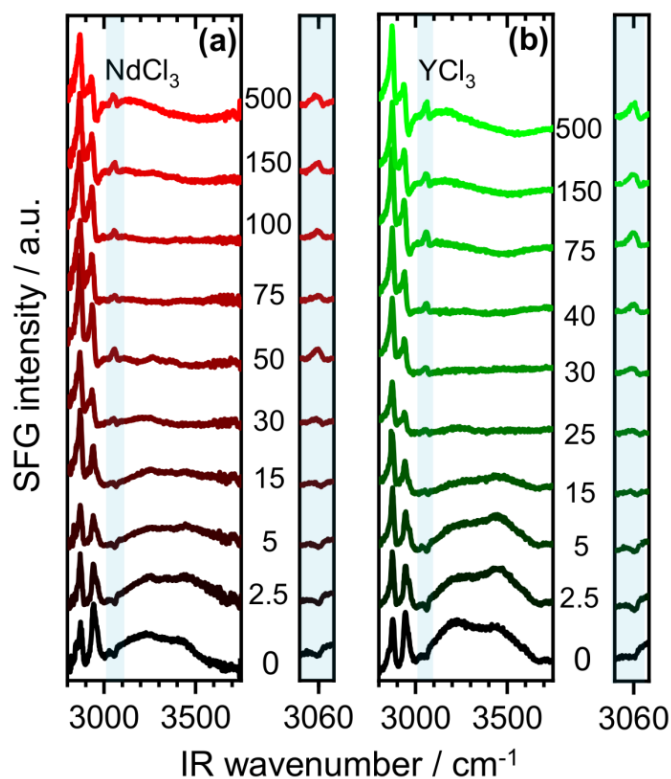

**Figure 2** *Vibrational SFG spectra of β-lactoglobulin (BLG) modified air-water interfaces from (a) YCl$_3$ and (b) NdCl$_3$ electrolyte solutions with 15 µM BLG. The salt concentrations in µM were as indicated in (a) and (b). The light blue shaded areas highlight the change of the shape of aromatic C-H stretching vibrations at 3060 cm$^{-1}$ as a function of NdCl$_3$ and YCl$_3$ concentration.*

Nd$^{3+}$ and Y$^{3+}$ ion concentrations cause screening of the double layer but, most importantly, the interfacial net charge is reduced due to specific ion binding to interfacial BLG molecules. This causes a dramatic reduction in the double-layer potential $\phi_0$ as both, the surface charge density and $\phi_0$, are connected and can be expressed by the Grahame equation.[54] A considerable decrease in interfacial net charge thus causes $\phi_0$ to decrease and leads to a substantial reduction in $\chi^{(3)}$ contributions to the SFG intensity (see equation (1)). Note that the pre-factor in equation (1) can also lead to changes in $\chi^{(3)}$ contributions at constant double-layer potential due to interference effects. However, in the latter case an increase in ionic strength would lead first to an increase in SFG intensity with a maximum at an ionic strength of ~ 1 mM and to a subsequent decrease as the double layer gets screened.[51] The fact, that we see a decrease in SFG intensity at much lower ion concentrations, a minimum and a subsequent increase in intensity, favors the idea that the interfacial net charge, and thus $\phi_0$, changes with the $Nd^{3+}$ and $Y^{3+}$ ion concentration.

Furthermore, a close analysis of the SFG spectra in Figure 2 also provides evidence for the charge reversal at the interface. Such interface overcharging induced by trivalent metal ions has already been reported for phospholipid monolayers.[55,56] For that, we now discuss the changes in the shape of the 3060 cm$^{-1}$ band due to aromatic C-H stretching vibrations (shaded areas in Figure 2). For both salts this bands appears as highly dispersive feature with a dip in SFG intensity close to 3060 cm$^{-1}$, at low ion concentrations. The appearance of the band at 3060 cm$^{-1}$ changes from dip to peak-like at ion concentrations of ~40 µM onwards and can be associated to a change in the net orientation of interfacial water molecules. This can be caused by a charge and consequently electric field reversal at the interface. As a result, polarization and net orientation of interfacial water is reversed, which induces a change in the polarity (phase) of O-H stretching vibrations (see details section). This effect is in fact similar to pH induced changes in SFG spectra when the interfacial isoelectric point is crossed.[16,57] As a result, constructive interference between the contributions from O-H and C-H stretching vibrations is observed when the $Nd^{3+}$ and $Y^{3+}$ ion concentration is high (peak-like feature at 3060 cm$^{-1}$, Figure 2), but is changed to destructive interference when the ion concentration is low (dip-like feature).

The recovery of the SFG intensity and polarity change in O-H bands for ion concentrations >40 µM corresponds to overcharging of the interface that is covered with positively charged BLG/ion complexes. This is also reflected in the overall shape of the O-H stretching bands in terms of relative intensities from the low and high frequency branches, which are dominated by the low frequency part of the spectrum at high ionic strengths while at low $Nd^{3+}$ and $Y^{3+}$ concentrations both, high and low frequency branches, contribute equally to the SFG spectrum (Figure 2).

In order to gain more detailed information on the interfacial point of zero net charge that is caused by binding of trivalent ions to the layer of interfacial BLG molecules, we determined the SFG intensity for each spectrum. In Figures 1b and 1c we provide a summary for the low and high frequency branches to the SFG intensity of the O-H spectrum, respectively. While the overall intensity is lower in the case of $Nd^{3+}$, the general trend of the intensity decrease and increase with ion concentration is not significantly different between $Nd^{3+}$ and $Y^{3+}$. For both ions, a minimum in O-H intensity of the low frequency branch is observed at ~40 µM while the high frequency branch of the O-H band just reaches a low plateau value. Compared to the bulk (Figure 1a), slightly higher concentrations of $Nd^{3+}$ and $Y^{3+}$ ions are required in order to reach charge neutral conditions at the air-water interface.

Obviously, binding of $Nd^{3+}$ and $Y^{3+}$ to BLG molecules in the bulk and at the air-water interfaces is very efficient and in both cases (interface vs. bulk) we presented information on the concentration dependent charging behavior. The questions on possible binding sites for $Nd^{3+}$ and $Y^{3+}$ ions was, however, not addressed so far. Previously, molecular dynamic simulations by Beierlein et al.[18] have shown that binding of $Li^+$ can occur via ion pair formation of $Li^+$ with Asp or Glu amino acids residues at the protein surface. For that reason, we studied carboxylate and Amid I stretching vibrations from interfacial BLG as a function of $Y^{3+}$ concentration. The results are presented in Figure 3, where a band at 1410 -1455 $cm^{-1}$ and a second band ~1655 $cm^{-1}$ dominate the SFG spectra. The former band is attributable to carboxylate symmetric stretching vibrations while the latter band can be assigned to so-called Amid I vibrations of carbonyl stretching modes from amino acids in the protein backbone.[16] The carboxylate band originates from Glu and Asp amino acid residues at the BLG surface.

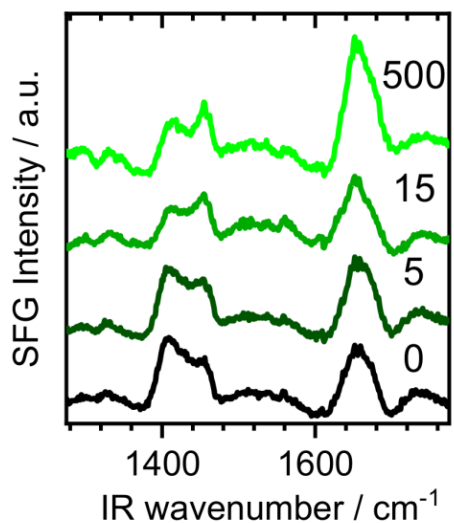

*Figure 3* Vibrational SFG spectra of β-lactoglobulin (BLG) modified air-water interfaces showing Amid I (1655 cm$^{-1}$) and carboxylate (-COO$^-$) stretching vibrations (1410 – 1450 cm$^{-1}$) as a function of YCl$_3$ concentration. The salt concentrations in µM were as indicated in the figure.

The Amid I band closely follows the change in surface coverage as can be qualitatively seen from the change in surface tension (Figure 1d). Although the intensity of the Amid I band is increased at 500 µM Y$^{3+}$ due to the increase in surface excess of BLG moieties at the air-water interface. The changes in Amide I band position and spectral line shape are, however, negligible. From this we conclude that there is no significant unfolding of secondary structure elements that give rise to SFG active Amide I contributions. While the Amide I band increases only in intensity, a spectral analysis of the carboxylate band provides more information on ion binding. The carboxylate band is split into two bands at 1410 and 1450 cm$^{-1}$, with a reversal in the intensity ratio of the latter two bands when the ion concentration is increased. At high Y$^{3+}$ concentrations, the relative contribution of the 1450 cm$^{-1}$ band increases and compared to the 1410 cm$^{-1}$ band which dominates this part of the SFG spectrum at 500 µM concentrations of Y$^{3+}$. From diverse previous works it is known that the frequency of carboxylate stretching bands strongly depends on the coordination with adjacent ions. This can cause substantial shifts in stretching frequency.[50,58–60] For instance, Tang et al.[58] reported in their SFG study on the interaction of $Ca^{2+}$ ions with carboxylate groups of palmitic acid layers at the air-water interface, a considerable blue-shift in frequency from 1435 to 1475 $cm^{-1}$ which the authors have assigned to a change in ion coordination from a 2:1 bridging complex to a 1:1 bidentate configuration. Different behavior was found by Tyrode and Corkery for arachidic acid in the presence of $Na^+$ cations, where the carboxylate stretching vibrations remained at 1408 $cm^{-1}$ for all concentrations. Interactions of arachidic acid with $La^{3+}$ ions introduced, however, a band at 1460 $cm^{-1}$.[61] From the above discussion on the frequency of carboxylate stretching frequencies from more ideal fatty acid monolayers at air-water interfaces and our own observation that the SFG intensity of the 1455 $cm^{-1}$ band increases on the costs of the intensity of the peak at 1410 $cm^{-1}$ when the $Y^{3+}$ concentration increases (Figure 3), we derive the following conclusion: $Y^{3+}$ ions bind to the carboxylate groups of interfacial BLG molecules and form ion pairs. However, at this point we also note that the SFG intensity at 1455 $cm^{-1}$ at zero salt concentration is for BLG moieties at the air-water interface non-zero and is likely associated to the $CH_3$ bending mode.[50] Therefore, at elevated salt concentration the 1455 $cm^{-1}$ is caused by the overlap of the carboxylate stretching mode (as discussed above) and the CH3 bending mode.

Considering previous work by Tang et al.[58], $Y^{3+}$ ions are likely bound in a 1:1 bidentate complex and binding is highly efficient. The latter two conclusions are corroborated by our analysis of the charging state in the bulk and by our analysis of O-H stretching bands, which indicate that charge neutral conditions are reached at $c_{ion}/c_p$ molar rations between 2 and 3. This indicates that on average between 2 and 3 $Y^{3+}$ or $Nd^{3+}$ ions are bound to one BLG molecule and compensate the BLGs net charge of ~-8e [36,62] at close to neutral pH.

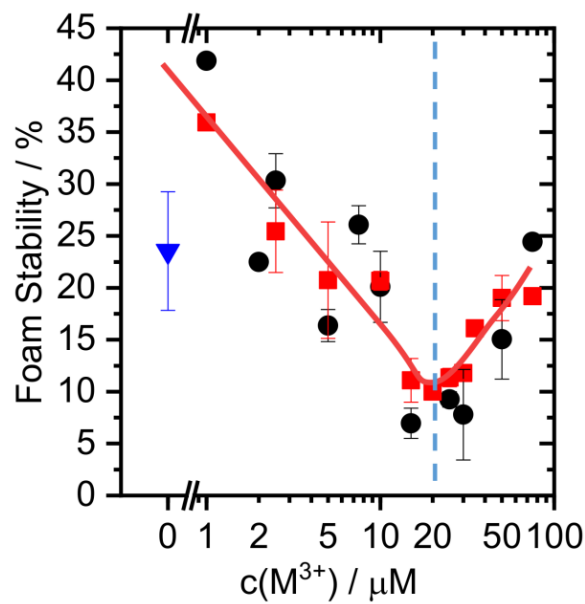

**Figure 4** *Foam stability (as defined in the details section) after 30 min for 15 µM BLG solutions as a function of NdCl$_3$ (red squares) and YCl$_3$ (black circles) electrolyte concentration. The data point marked with a blue triangle represents the stability of foams from pure 15 µM BLG solutions without salt. Solid and dashed lines guide the eye.*

**Effects of surface and bulk charging state on foam stability**

In Figure 4, we present the foam stability as defined in the experimental details section. Here, we studied aqueous foams from 15 µM BLG solutions with and without addition of NdCl$_3$ and YCl$_3$. As it can be inferred from Figure 4, the foam stability first decreases with rising ion concentration and reaches a minimum at ~20 µM for both cations. Further increase of the NdCl$_3$ or YCl$_3$ concentration from 20 to 100 µM leads to a substantial re-increase in the foam stability.

By comparing Figure 1a and Figure 1b with Figure 4, it becomes obvious that the changes in the charging condition of BLG molecules in the bulk solution and at the air-water interface for different NdCl$_3$ and YCl$_3$ concentrations are also reflected in the stability of aqueous foams from such solutions. These results, which point to a relation between interfacial net charge and foam stability, can be explained with the classical DLVO (Derjaguin, Landau, Verwey, Overbeek) theory. This theory describes the stability of foam films (lamellae) by a balance of repulsive electrostatic and attractive dispersive interactions.[54] In ideal systems, the disjoining pressure of a single foam film can be calculated for a given double layer potential $\phi_0$ with the help of the DLVO theory and it can account for bubble coalescence and the stability of the macroscopic foam when other degradation mechanisms are less dominant. In particular, the electrostatic disjoining pressure, that depends on $\phi_0$, can be qualitatively inferred from the change in the SFG intensity of O-H stretching bands (Figure 1b) which also depends on $\phi_0$ via third-order $\chi^{(3)}$ contributions (equation (1)). A reduction of the protein net charge leads to a reduced interfacial net charge (Figure 1b),[18,23] and consequently to a drop of the electrostatic disjoining pressure. This causes a decrease of the stability of the macroscopic foam. At the point of zero net charge in the bulk and at the interface, which are at concentrations of 30 and 40 µM, respectively, the foam stability is poor. This is consistent with our conclusion above that the change of the electrostatic disjoining pressure as a function of the NdCl$_3$ and YCl$_3$ concentration drives the changes in foam stability (Figure 4). However, so far we addressed only the charging effects at the ubiquitous air-water interfaces, while in the bulk also protein aggregates are formed in regime II between the characteristic concentrations of $c_i^*$~15-30 µM and $c_i^{**}$~120-300 µM (Figure 1a, Figure S1). Consequently, we need to discuss the impact of BLG aggregates on the foam stability within regime II. Previously, Rullier et al.[63] concluded that the properties and stability of BLG foam films and the corresponding macroscopic foams can be either improved or disrupted depending on the size of BLG aggregates and the amount of non-aggregated BLG. The authors found that in mixtures of 96 % thermally aggregated BLG molecules and 4 % non-aggregated BLG the foam stability was increased. For that reason, we propose that the apparent differences between the minimum in foam stability (Figure 4) at 20 µM concentration and the point of zero net charge in the bulk (~30 µM, Figure 1a) and at the interface (~40 µM, Figure 1b) are associated to the presence of aggregates which lead to a non-DLVO type foam stabilization similar to Pickering foam.[64] Pickering stabilization may compensate the loss in electrostatic disjoining pressure at concentrations where the interfacial net charge is low. Increasing the $Y^{3+}$ and $Nd^{3+}$ concentration above 30 µM induces an excess of positive charges at the BLG molecules (Figure 1a) and causes an overcharging in the bulk and at the air-water interface (Figure 1b). Consequently, this leads to an increase of the electrostatic disjoining pressure and to higher foam stabilities even in the absence of BLG aggregates.

**SUMMARY AND CONCLUSIONS**

We studied the influence of trivalent cations $Y^{3+}$ and $Nd^{3+}$ on the bulk, interfacial and foaming properties of β-lactoglobulin (BLG) molecules in aqueous solutions using a multi-technique approach. Binding of the latter trivalent ions to BLG molecules causes a remarkable reduction of the protein net charge and a charge reversal which occurs between 30 to 40 µM $Y^{3+}$ or $Nd^{3+}$ ($c_i/c_p \approx 2 - 3$) concentrations. This reduction in the protein net charge is accompanied by substantial aggregation and sedimentation of BLG molecules in the bulk solution, which was confirmed by turbidity measurements. The concentration range of this condensed regime is estimated between 15 and 300 µM. The charging state and the binding of $Y^{3+}$ and $Nd^{3+}$ ions to BLG molecules at the air-water interface is studied in detail with surface-specific SFG spectroscopy. SFG shows a dramatic loss in O-H intensity until a minimum at ~40 µM ion concentration is reached that is followed by a subsequent rise in intensity. This behavior and the minimum in O-H intensity is attributed to a charge reversal at the interface with a point of zero net charge at ~40 µM of $Y^{3+}$ or $Nd^{3+}$. This conclusion is corroborated by a phase analysis of O-H stretching vibrations, which shows that the waters' net orientation at the interface is reversed at an ion concentration of ~40 µM. It is interesting to note that there are no significant differences in the binding behavior for $Y^{3+}$ or $Nd^{3+}$ cations. However, in comparison to alkali or $Ca^{2+}$ cations, binding of $Y^{3+}$ or $Nd^{3+}$ is orders of magnitude more efficient. In order to address possible binding sites of the $Y^{3+}$ or $Nd^{3+}$ ions at BLG moieties, we recorded SFG spectra of Amide I and carboxylate stretching vibrations and provided evidence for ion binding at carboxylate groups from Asp or Glu residues from interfacial BLG molecules. The charging behavior in the bulk and at air-water interfaces, as well as bulk aggregation, is shown to drive the foam stability through structure-property relations, with a minimum in foam stability close to the interfacial and bulk point of zero net charge.

## ASSOCIATED CONTENT

Supporting Information on a time dependent analysis of BLG aggregation and sedimentation behavior and a comparison of dynamic surface tensions at short and long time periods is available.

## ACKNOWLEDGMENT

We would like to thank the group of Ulrich Kulozik (Technical University of Munich, Germany) for providing us with high quality β-lactoglobulin. We are thankful for the experimental assistance from Caroline Mönich and Michael Groh. Additionally we are grateful for the funding of the European Research Council (ERC) under the European Union's Horizon 2020 research and innovation program (grant agreement No 638278) and the Deutsche Forschungsgemeinschaft (DFG).

**For Table of Contents Only**

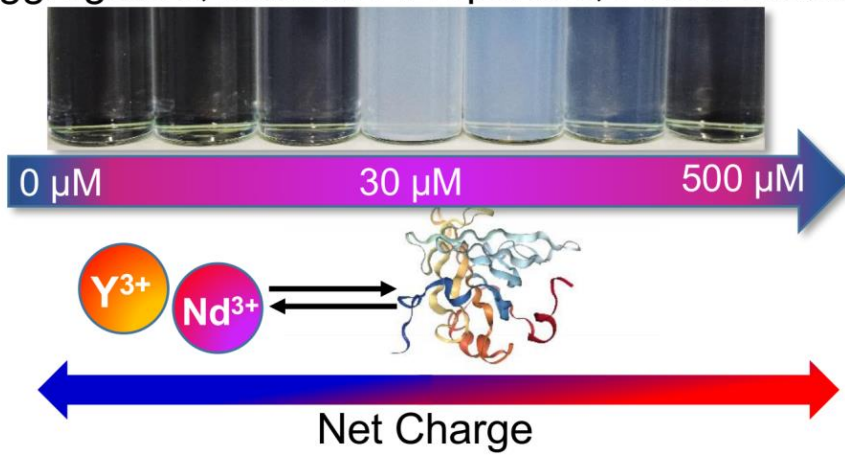